\documentclass[conference]{IEEEtran}
\usepackage{fancyhdr}
\IEEEoverridecommandlockouts
\usepackage{algorithm}
\usepackage{algpseudocode}
\usepackage{amsfonts}
\usepackage{bm}
\usepackage{amsmath}
\usepackage{cite}
\usepackage{amssymb}
\usepackage{enumerate}
\usepackage{color,xcolor}
\usepackage{graphicx}
\usepackage{multirow,tabularx}
\usepackage{subfigure}
\usepackage{hyperref}

\makeatletter

\newcommand{\Rmnum}[1]{\expandafter\@slowromancap\romannumeral #1@}
\makeatother

\ifCLASSINFOpdf

\else

\fi

\PassOptionsToPackage{draft}{hyperref}

\begin{document}

\title{{Capacity-Achieving Iterative LMMSE Detection for MIMO-NOMA Systems}}

\author{\IEEEauthorblockN{Lei Liu\IEEEauthorrefmark{1}, Chau Yuen\IEEEauthorrefmark{2}, Yong Liang Guan \IEEEauthorrefmark{3} and Ying Li\IEEEauthorrefmark{1} \\
\IEEEauthorrefmark{1}State Key Lab of ISN, Xidian University, Xi'an 710071, China\\
\IEEEauthorrefmark{2}Singapore University of Technology and Design, Singapore\\
\IEEEauthorrefmark{3}Nanyang Technological University, Singapore\\
E-mail:yli@mail.xidian.edu.cn
}

\thanks{This work was supported in part by the 973 Program under Grant 2012CB316100, the National Natural Science Foundation of China under Grants 61301177 and 61550110244, and Singapore A*STAR SERC Project under Grant 142 02 00043. The first author was also supported by the China Scholarship Council under Grant 20140690045.}}


\maketitle

\begin{abstract}
This paper considers a iterative Linear Minimum Mean Square Error (LMMSE) detection for the uplink Multiuser Multiple-Input and Multiple-Output (MU-MIMO) systems with Non-Orthogonal Multiple Access (NOMA). The iterative LMMSE detection greatly reduces the system computational complexity by departing the overall processing into many low-complexity distributed calculations. However, it is generally considered to be sub-optimal and achieves relatively poor performance. In this paper, we firstly present the matching conditions and area theorems for the iterative detection of the MIMO-NOMA systems. Based on the proposed matching conditions and area theorems, the achievable rate region of the iterative LMMSE detection is analysed. We prove that by properly design the iterative LMMSE detection, it can achieve \emph{(i)} the optimal sum capacity of MU-MIMO systems, \emph{(ii)} all the maximal extreme points in the capacity region of MU-MIMO system, and \emph{(iii)} the whole capacity region of two-user MIMO systems.

\end{abstract}


\IEEEpeerreviewmaketitle
\section{Introduction}
Recent research investigations\cite{5GWhitepaper, Argas2013,Rusek2013,biglieri2007,Yang2014} show that Multiuser Multiple-Input and Multiple-Output (MU-MIMO) will play a vital role in the fifth generation (5G) mobile networks. MU-MIMO has become a key technology for wireless communication standards like IEEE 802.11 (Wi-Fi), WiMAX (4G) and Long Term Evolution (4G). Especially, the massive MU-MIMO has attracted a lot of attentions \cite{Argas2013, Rusek2013,biglieri2007, Ngo2012, Dai2013, Han_acpt} because of its improvement both in throughput and energy efficiency \cite{ Ngo2012, Dai2013,Han_acpt}. In addition, Non-Orthogonal Multiple Access (NOMA) has also been identified as one of the key radio access technologies to further increase system spectral efficiency and reduce latency in the 5G communictions systems \cite{5GWhitepaper, Saito2013, Al-Imari2014, Ding2014,Lei1,Lei2}.

The costs are more physical space at Base Station (BS), higher complexity, and higher energy consumption of the signal processing at both ends\cite{Rusek2013,biglieri2007}. Low-complexity signal uplink detection for MIMO-NOMA is one of these current challenging problems\cite{Rusek2013}. The optimal multiuser detector (MUD), such as the maximum a posterior probability (MAP) detector or maximum likelihood (ML) detector, was proven to be an NP-hard and non-deterministic polynomial-time complete (NP-complete) problem \cite{Micciancio2001,verdu1984_1}. Thus, the complexity of optimal MUD grows exponentially with the number of users or the number of antennas at the BS and polynomially with the size of the signal constellation \cite{verdu1987}. Many low-complexity linear detections such as Matched Filter, Zero-Forcing receiver \cite{tse2005}, Minimum Mean Square Error (MMSE) detector and Message Passing Detector (MPD) \cite{Loeliger2004, Loeliger2006} are proposed for the practical systems. Although these linear MUDs are attractive from the complexity view point, they achieve relatively poor performance for MU-MIMO systems.

The iterative detections that exchange soft information of the low-complexity detector with the user decoders are mostly used as an efficient receivers for practical MIMO-NOMA systems \cite{andrea2005, Gao2014, Lei2015}. This is a fundamental technology for the NOMA like the Code Division Multiple Access (CDMA) \cite{tse2005,verdu1998} and the Interleave Division Multiple Access IDMA systems \cite{Ping2003_1}. Various iterative detectors, such as the iterative Linear MMSE (LMMSE) detector, iterative BP detector and iterative MPD, were proposed to achieve a good system performance \cite{Guo2008, Caire2004, Yuan2014}. The iterative detection is a low-complexity parallel joint iterative decoding method, which further reduces the detection complexity by departing the overall receiver into many distributed processors. However, in general, the joint iterative detection structure cannot achieve the perfect performance and is considered to be sub-optimal \cite{verdu1998}. Therefore, the achievable rate region of the MIMO-NOMA systems with iterative detection is an intriguing problem.

The Extrinsic Information Transfer (EXIT) \cite{Ashikhmin2004,Brink2001}, MSE-based Transfer Chart (MBTC) \cite{Guo2005,Bhattad2007,Yuan2014}, area theorem and matching theorem \cite{Ashikhmin2004, Brink2001, Bhattad2007, Yuan2014, Guo2005} are the main methods of the system achievable rate or the BER performance analysis. It is proved that a well-designed single-code with linear precoding and iterative LMMSE detection achieves the capacity of the MIMO systems \cite{Yuan2014}. In this paper, we consider a low-complexity iterative LMMSE detection for the uplink MIMO systems with NOMA. The achievable rate analysis of the iterative LMMSE detection is provided, which shows it is rate region optimal for the MU-MIMO systems if properly designed. The contributions of this paper are listed as follows:

\noindent
\hangafter=1
\setlength{\hangindent}{2em} 1) For the MU-MIMO systems, matching conditions and area theorems for iterative detection are proposed.

 \noindent
\hangafter=1
\setlength{\hangindent}{2em} 2) With the matching conditions and area theorems, the iterative LMMSE detection design and its achievable rate analysis for the MIMO-NOMA systems are provided.

 \noindent
\hangafter=1
\setlength{\hangindent}{2em} 3) We prove that the designed iterative LMMSE detection \emph{(i)} is sum capacity achieving for the MU-MIMO systems, \emph{(ii)} achieves all the maximal extreme points in the capacity region of MU-MIMO system, and \emph{(iii)} achieves the whole capacity region of two-user MIMO systems.

This paper is organized as follows. In Section II, the MIMO-NOMA system model is introduced. The matching conditions and area theorems for the MU-MIMO systems are elaborated in Section III. Section IV provides the achievable rate region analysis for the MIMO-NOMA systems with iterative LMMSE detection. Some special cases are shown in Section V, and we end with conclusions in Section VI.

\section{System Model}
\begin{figure}[t]
  \centering
  \includegraphics[width=9cm]{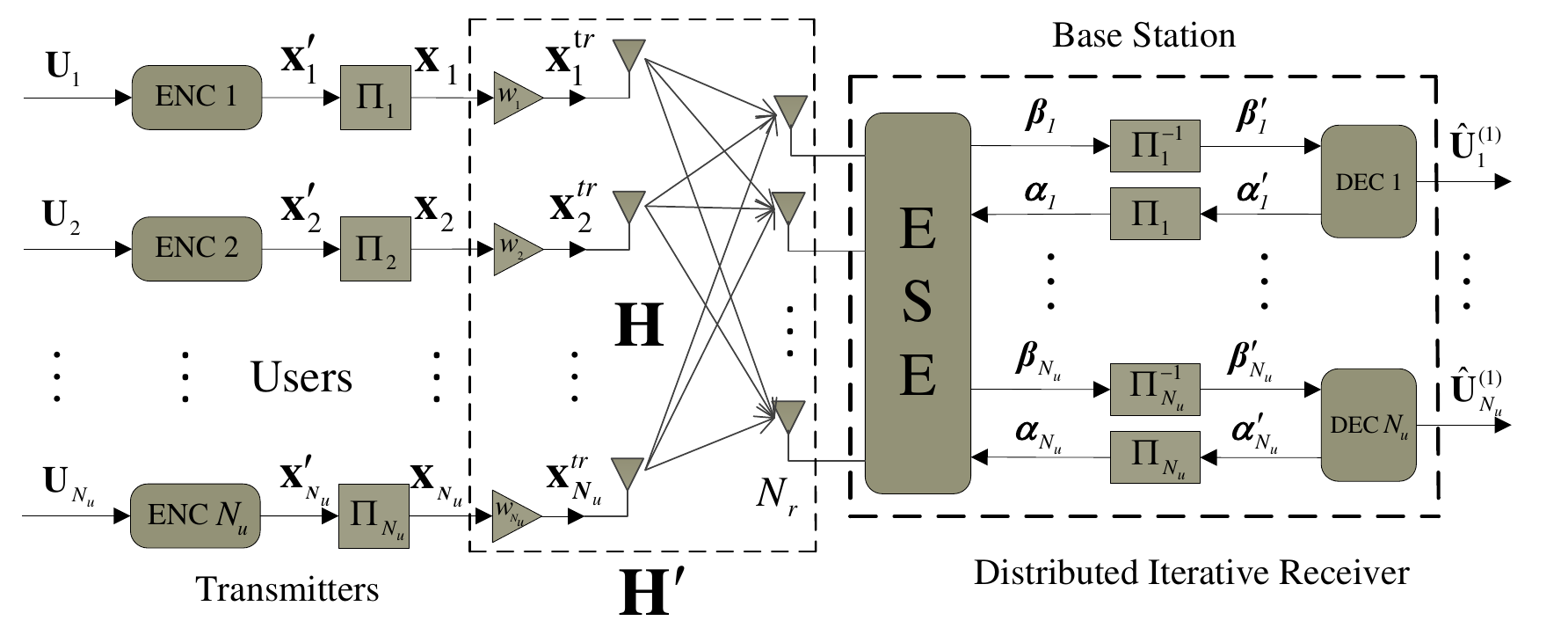}\\
  \caption{Block diagram of the multiuser MIMO system. ENC is the encoder and DEC is the decoder. $\Pi_i$ denotes the interleaver and $\Pi_i^{-1}$ denotes the de-interleaver. ESE represents the elementary signal estimator. $\mathbf{H}'$ contains the channel $\mathbf{H}$ and $K_{\mathbf{x}}=diag\{w_1^2,\cdots,w_{N_u}^2\}$, and $K_{\mathbf{x}}$ denotes the power constraint or the large scale fading of the users.  }\label{f2}\vspace{-0.3cm}
\end{figure}

Consider a uplink MU-MIMO system with NOMA as showed in Fig \ref{f2}.  In this system, $N_{u}$ autonomous single-antenna terminals simultaneously communicate with an array of $N_{r}$ antennas of the base station (BS) in the same frequence and at the same time\cite{Ngo2012,Rusek2013}. At user $i$, an information sequence ${\bf{U}}_i $ is encoded by a channel code with rate $R_i$ into a $N$-length coded sequence ${\mathbf{x}}'_{i}$, $\mathop{i}\in \mathcal{N}_{u},\; \mathcal{N}_{u}= \left\{ {{{1,2,}} \cdots {{,N_{u}}}} \right\}$ and then interleaved by an {$N$-length} independent random interleaver $\Pi_{i}$ and get $\mathbf{x}_{i}=[x_{i,1},x_{i,2},\cdots,x_{i,N}]^T$. We assume that $x_{i,t}$ is randomly and uniformly taken over the points in a discrete signaling constellation $\mathcal{S}=\{s_1,s_2,\cdots,s_{|\mathcal{S}|}\}$. After that, the $\mathbf{x}_{i}$ is scaled with $w_i$, which denotes the power constraint or the large-scale fading coefficient of each user, and we then get the transmitting $\mathbf{x}^{tr}_{i}$, $\mathop{i}\in \mathcal{N}_{u}$. 

Then the $N_{r}\times1$ received signal vector ${ \mathbf{y}_t}$ at time $t$ is
\begin{eqnarray}\label{e1}
{ \mathbf{y}_t}=\mathbf{H}{ K_{\mathbf{x}}^{1/2}\mathbf{x}(t)} + \mathbf{n}(t)
= \mathbf{H}'{\mathbf{x}(t)} + \mathbf{n}(t),\;\; t\in \mathcal{N}
\end{eqnarray}
where $\mathcal{N}=\{1,\cdots,N\}$, $\mathbf{H}$ is a given $N_{r} \times N_{u}$ channel matrix, $\mathbf{n}(t)\sim\mathcal{CN}^{{N_{r}}}(0,\sigma_n^2)$ is the $N_{r} \times 1$ Gaussian noise vector at time $t$, and $\mathbf{x}^{tr}(t)=[x^{tr}_1(t),\cdots,x^{tr}_{N_u}(t)]^T$ is the message vector from $N_{u}$ users.

At the BS, the received $\mathbf{Y}=[\mathbf{y}_1,\cdots,\mathbf{y}_N]$ and message $\bm{\alpha}_i$ from the decoder are sent to a low-complexity elementary signal estimator (ESE) to estimate the extrinsic message $\bm{\beta}_i$, which is then deinterleaved with $\Pi_i^{-1}$ into $\bm{\beta}'_i$. The corresponding single-user decoder employs $\bm{\beta}'_i$ as the prior message to calculate the extrinsic message $\bm{\alpha}'_i$. Similarly, this extrinsic message is interleaved by $\Pi_i$ to obtain the prior information $\bm{\alpha}_i$ for the ESE. Repeat this process until the maximum number of iteration is achieved or the messages are recovered correctly.

\section{Matching Conditions and Area Theorems for MU-MIMO Systems}
In this section, we proposed the matching conditions and area properties for the iterative MU-MIMO systems from their \emph{SINR-Variance} transfer functions. It should be noted that the results in this section are generalized based on the \emph{I-MMSE} theorem and the area theorems that proposed for the single user MIMO systems\cite{Guo2005,Bhattad2007,Yuan2014}.

\subsection{Characterization of ESE}
We first define the \emph{SINR-Variance} transfer function of user $i$ of element signal estimator as
\begin{equation}\label{ese1}
\phi_i(\mathbf{v}_{\bar{\mathbf{x}}})=v_{\hat{x}_i}^{-1}-v_i^{-1},\;\; \mathrm{for}\;\; i\in \mathcal{N}_u,
\end{equation}
where $\mathbf{v}_{\bar{\mathbf{x}}}=[v_1,\cdots,v_{N_u}]$, $v_{\hat{x}_i}$ is the $i$th diagonal element of covariance matrix $\mathbf{V} _{{{\hat {\mathbf{x}}}}}$, $v_i$ is the input variance ($\bm{\alpha}_i$) of user $i$, i.e., the $i$th diagonal element of $\mathbf{V} _{{{ \bar{\mathbf{x}}}}}$, and $\phi_i(\mathbf{v})$ denotes the output extrinsic \emph{SINR} of user $i$ at the estimator. The variance $v_i$ varies from $0$ to $1$ as the signal power is normalized to 1. Similarly, the total MSE of user $i$ at the estimator is $\mathrm{mmse}_{tot,i}^{ese}(\mathbf{v}_{\bar{\mathbf{x}}}) = v_{\hat{x}_i}$. The next Gaussian assumption is used to simplify the system analysis, which is a common assumption in many works \cite{Kay1993,Poor1997}.

\emph{Assumption 1: Let $\bm{\rho}=[\rho_1,\cdots,\rho_{N_u}]$, $\bm{\phi}(\mathbf{v}_{\bar{\mathbf{x}}})= \left[{\phi_1}(\mathbf{v}_{\bar{\mathbf{x}}}),\cdots, {\phi_{N_u}}(\mathbf{v}_{\bar{\mathbf{x}}}) \right]$. The outputs $[\bm{\beta}_1,\cdots,\bm{\beta}_{N_u}]$ of the estimator can be approximated as the observations from AWGN channels and the related \emph{SINR} is denoted by $\bm{\rho}=\bm{\phi}(\mathbf{v}_{\bar{\mathbf{x}}})$. }

\subsection{Characterization of APP Decoder}
From Assumption 1, the input of the each decoder $\bm{\beta}'_i$ are equivalent as the independent observations over an AWGN channel with $SNR_i=\rho_i$. For any $i\in \mathcal{N}_u$, we define the \emph{SINR-Variance} transfer function of the decoder as
\begin{equation}\label{e24}
v_i=\psi_i(\rho_i).
\end{equation}
 Let $\bm{\psi}(\bm{\rho})=[\psi_1(\rho_1),\cdots,\psi_{N_u}(\rho_{N_u})]$, and we get
$\mathbf{v}_{\bar{\mathbf{x}}}=\bm{\psi}(\bm{\rho})$.

\subsection{SINR-Variance Transfer Chart}
The LMMSE estimator is described by $\bm{\rho}=\bm{\phi}(\mathbf{v}_{\bar{\mathbf{x}}})$, and the decoders can be described by $\mathbf{v}_{\bar{\mathbf{x}}}=\bm{\psi}(\bm{\rho})$. Therefore, the iterative detection performs iteration between the estimator and the decoders and can be tracked by the values of $\bm{\rho}$ and $\mathbf{v}_{\bar{\mathbf{x}}}$.
The estimator and decoders are matched if
\begin{equation}\label{e29}
\bm{\phi}\left(\mathbf{v}_{\bar{\mathbf{x}}}\right) =\bm{\psi}^{-1}\left(\mathbf{v}_{\bar{\mathbf{x}}}\right),\quad \mathrm{for} \;\; \mathbf{0}  < \mathbf{v}_{\bar{\mathbf{x}}} \leq \mathbf{1}.
\end{equation}
It means that $\phi_i(\mathbf{v}_{\bar{\mathbf{x}}})=\psi_i^{-1}(v_i)$ for any $i\in \mathcal{N}_u$. The matched transfer function not only maximizes the rate of the code, but also ensures the transmitting signals can be perfectly recovered. Note that: $\phi_i(\mathbf{1})>0$ as the estimator always use the information from the channel; and $\phi_i(\mathbf{0})>1$ as the estimator cannot remove the uncertainty introduced by the channel noise. Therefore, we have the following proposition.

\emph{Proposition 1}: \emph{For any $i\in \mathcal{N}_u$, the matching conditions of the iterative MU-MIMO systems can be rewritten as}
\begin{eqnarray}
\psi_i(\rho_i)&=&\phi_i^{-1}(\phi_i(\mathbf{1}))=1, \;\;\mathrm{for}\;\; 0\leq\rho_i<\phi_i(\mathbf{1});\label{e30}\\
\psi_i(\rho_i)&=&\phi_i^{-1}(\rho_i),\;\;\mathrm{for}\;\; \phi_i(\mathbf{1})\leq\rho_i<\phi_i(\mathbf{0});\label{e31}\\
\psi_i(\rho_i)&=&0,\;\;\mathrm{for}\;\; \phi_i(\mathbf{0})\leq\rho_i<\infty.\label{e32}
\end{eqnarray}
Similarly, ${\phi}_i^{-1}(\cdot)$ denotes the inverse of ${\phi}_i(\cdot)$. 

\subsection{Area Properties}
 Let $\mathrm{snr}_{ap,i}^{dec}$ denote the \emph{SNR} of the prior input messages in decoder $i$, $\mathrm{snr}_{ext,i}^{ese}$ be the \emph{SNR} of the output extrinsic messages in estimator to the decoder $i$, $\mathrm{mmse}_{tot,i}^{dec}(\cdot)$ represent the total variance of the messages for user $i$ at the LMMSE estimator, and $\mathrm{mmse}_{tot,i}^{dec}(\cdot)$ indicate the total variance of the messages in decoder $i$.
In addition, $\mathbf{snr}_{ext,i}^{ese}=[\mathrm{snr}_{ext,1}^{ese}, \cdots,\mathrm{snr}_{est,N_u}^{ese}]$. The area properties is given as follows.

 \emph{Proposition 2}: \emph{The achievable rate $R_i$ of user $i$ and an upper bound of $R_i$ are given as}
 \vspace{-0.25cm}
 \begin{equation}\label{e34}
\!\!\!\!{R_i} \!\!=\!\!\!\! \int\limits_0^\infty  \!\!\! {\mathrm{mmse}_{tot\!,i}^{dec}\!(\!{{snr}}_{\!\!ap\!,i}^{\!dec}\!)d} \mathrm{snr}_{\!ap\!,i}^{dec},
R_i^{\max } \!\!=\!\!\!\! \int\limits_0^\infty  \!\!\! {\mathrm{mmse}_{\!tot\!,i}^{ese}(\!\mathbf{snr}_{ext}^{ese}\!)d} \mathrm{snr}_{\!ext\!,i}^{ese},
\vspace{-0.15cm}
\end{equation}
\emph{and $R_i\leq R_i^{\mathrm{max}}$, $i\in \mathcal{N}_u$, where the equality holds if and only if the \emph{SINR-Variance} transfer functions of the element signal estimator and decoders for any user are matched with each other, i.e., (\ref{e29}) and the matching conditions (\ref{e30})$\sim$ (\ref{e32}) hold.}

In our MU-MIMO system model, from (\ref{ese1}) and (\ref{e24}) and with the Gaussian assumptions , we have $\mathrm{snr}_{ap,i}^{dec}=\rho_i$, $\mathrm{snr}_{ext,i}^{ese}=\phi_i(\mathbf{v}_{\bar{\mathbf{x}}})$, $\mathrm{mmse}_{tot,i}^{dec}(\mathbf{{snr}}_{ap}^{dec})={{\left( {{\rho _i} + {\psi _i}{{({\rho _i})}^{ - 1}}} \right)}^{ - 1}}$ and $\mathrm{mmse}_{tot,i}^{ese}(\mathbf{snr}_{ext,i}^{ese})=v_{\hat{x}_i}(\mathbf{v}_{\bar{\mathbf{x}}})$.
Therefore, (\ref{e34}) can be rewritten as the following proposition.

\emph{Proposition 3}:\emph{ With the \emph{SINR-Variance} transfer functions $\bm{\rho}=\bm{\phi}(\mathbf{v}_{\bar{\mathbf{x}}})$ and $\mathbf{v}_{\bar{\mathbf{x}}}=\bm{\psi}(\bm{\rho})$ and the Gaussian assumptions,} the achievable rate $R_i$ of user $i$ and an upper bound of $R_i$ are\vspace{-0.2cm}
 \begin{eqnarray}
\!{R_i} = \!\!\int\limits_0^\infty \!\!\! {{{\left( {{\rho _i} + {\psi _i}{{({\rho _i})}^{ - 1}}} \right)}^{ - 1}}d{\rho _i}}\label{e35},\;
R_i^{\max }\!\! =\!\!\! \int\limits_0^\infty \!\! v_{\hat{x}_i}(\mathbf{v}_{\bar{\mathbf{x}}}) d \phi_i(\mathbf{v}_{\bar{\mathbf{x}}}).\label{e36}
\end{eqnarray}\vspace{-0.1cm}
\emph{and $R_i\leq R_i^{\mathrm{max}}$, $i\in \mathcal{N}_u$, where the equality holds if and only if the \emph{SINR-Variance} transfer functions of the element signal estimator and decoders for any user are matched with each other£¬ i.e., (\ref{e29}) and the matching conditions (\ref{e30})$\sim$ (\ref{e32}) hold.}

All the users' transfer functions interact with each other at the estimator since every output of the estimator depends on the variances of the input messages from all the decoders. In addition, all the users' transfer functions are unknown and need to be properly designed. Therefore, it is very hard to calculate the achievable rates directly with (\ref{e36}).

\section{Achievable Rate Region Analysis of Iterative LMMSE Detection}
In this section, based on the proposed matching conditions and area properties, the achievable rates of users are given for MIMO-NOMA systems with iterative LMMSE detection. In the iterative LMMSE detection, we use the LMMSE estimator as the ESE estimator and the Superposition Code Modulation (SCM) codes as the channel codes.

\subsection{LMMSE ESE}
LMMSE is an alternative low complexity ESE. Let $\bar{\mathbf{x}}(t)=[{x_{1,t}},\cdots,x_{N_u,t}]$ and $\mathbf{V}_{\bar{\mathbf{x}}(t)}=\mathbf{V}_{\bar{\mathbf{x}}}= \mathrm{diag}\{v_1,v_2,\cdots,v_{N_u}\}$. The LMMSE detector \cite{tse2005} is
\begin{eqnarray}\label{GMP2}
\!\!\!\!\!\!\!\!\!\!\!\!{{\hat {\mathbf{x}}}(t)}
\!\!\!\!\!&=&\!\!\!\!\! \left(\sigma_n^{-2}\mathbf{H}'^H\mathbf{H}'\!+\! \mathbf{V}_{\bar{\mathbf{x}}}^{-1}\right)^{-1}\!\!\left[\mathbf{V}_{\bar{\mathbf{x}}}^{-1}\bar{\mathbf{x}}(t)\!+\! \sigma_n^{-2}\mathbf{H}'^H\mathbf{y}_t  \right]\\
&=& \!\!\!\!\bar{\mathbf{x}}(t)\! +\! V_{\bar{\mathbf{x}}}\mathbf{H}'^H\!\left(\sigma_n^2\mathbf{I}_{N_r}\!\!\!+\! \mathbf{H}'V_{\bar{\mathbf{x}}}\mathbf{H}'^H \right)^{-1}\!\!\left(\mathbf{y}_t\!-\!\mathbf{H}'\bar{\mathbf{x}}(t)\right)\nonumber
\end{eqnarray}
where $\mathbf{V} _{{{\hat {\mathbf{x}}}}} = (\sigma _{{{n}}}^{- 2}\mathbf{H}'^H\mathbf{H}'+\mathbf{V} _{{{ \bar{\mathbf{x}}}}}^{-1})^{-1}$, which denotes the deviation of the estimation to the initial sources.

Therefore, with GMP2, we get $\beta_{i,t} = x_{i,t} + n^{*}_{i,t}$, and
\begin{equation}\label{e22}
\!n_{i,t}^{*}\!\!=\!\!  \frac{v_i}{v_{\hat{x}_i}\!\rho_i}\!v_{\bar{x}_i}^2\!{\mathbf{h}'_i}^H\!\!\!\left( \! \sigma^2_n\mathbf{I}_{N_r}\!\!\!\!+\!\! \mathbf{H}'\mathbf{V}_{\bar{\mathbf{x}}}\!\mathbf{H}'^H \!\right)^{-1}\!\!\!\left[\! \mathbf{H}'\!\!\left(\!\mathbf{x}_{\backslash i}(t)\!-\!\bar{\mathbf{x}}_{\backslash i}(t)\!\right) \!\!+\!\!\mathbf{n}(t)\!\right]
\end{equation}
where $\mathbf{x}_{\backslash i}(t)$ (or $\bar{\mathbf{x}}_{\backslash i}(t)$) denotes the vector whose $i$th entry of $\mathbf{x}(t)$ (or $\bar{\mathbf{x}}(t)$)  is zero. We rewrite Assumption 1 as follows.

\emph{Assumption 2: The equivalent noise $n_{i,t}^{*}$ is independent of $x_{i,t}$ and is Gaussian distributed $n_{i,t}\sim \mathcal{CN}\left(0,1/\phi_i(\mathbf{v}_{\bar{\mathbf{x}}})\right)$, i.e., the output of the LMMSE estimator is the observation from AWGN channel, i.e., $\bm{\beta}(t)=\mathbf{x}(t)+\mathbf{n}_t^{*}$ with SNRs $\bm{\rho}=\bm{\phi}(\mathbf{v}_{\bar{\mathbf{x}}})$.}

\subsection{A Posteriori Probability Decoders: SCM Decoder}
As the SCM code is capacity-achieving and easily analyzed \cite{Wachsmann1999,Gadkari1999}, a property is established in \cite{Yuan2014} and the area theorems \cite{Guo2005,Bhattad2007}, which builds the relationship between the rate of the SCM code and its transfer function $\psi_i(\rho_i)$.

\emph{Property of SCM Codes}: \emph{There exists such an $n$-layer SCM code $\Gamma_n$ whose transfer function can approach the function $\psi(\rho)$ with arbitrary small error if $n$ is large enough and $\psi(\rho)$ satisfies the following conditions:\\
(i) $\psi(0)=1$ and $\psi(\rho)\geq 0$, for $\rho\in[0,\infty);$\\
(ii) monotonically decreasing in $\rho\in[0,\infty)$;\\
(iii) continuous and differentiable in $[0,\infty)$ except for a countable set of values of $\rho$;\\
(iv) $\mathop {\lim }\limits_{{\rho } \to \infty } {\rho }{\psi }({\rho }) = 0$.}

\subsection{Sum Capacity Achieving of Iterative LMMSE Detection for MU-MIMO Systems}

\begin{figure}[t]
  \centering
  \includegraphics[width=7cm]{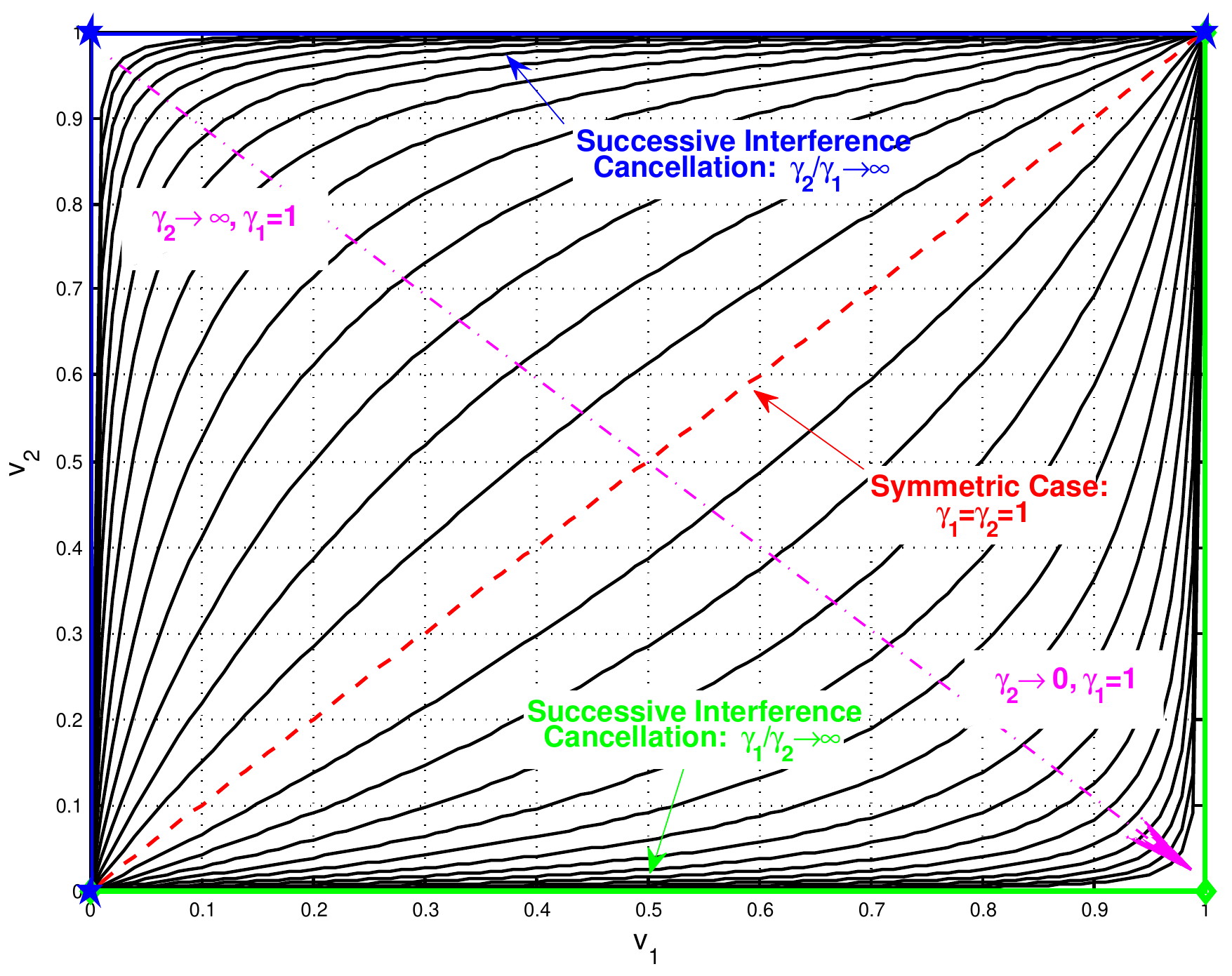}\\\vspace{-0.2cm}
  \caption{Variance tracks for the different $\bm{\gamma}$, where $\gamma_1=1$ is fixed. The variance of user $i$ is denoted as $v_i$, $i=1,2$. When $\gamma_2$ changes from $\infty$ to $0$, the variance track change from the blue curve (SIC case and the decoding order is user $1\to$ user $2$) to green curve (SIC case and the decoding order is user $2\to$ user $1$). When $\gamma_1=\gamma_2=1$, it is degenerated to the symmetric case (red line). }\label{f3}\vspace{-0.3cm}
\end{figure}

\begin{figure}[t]
  \centering
  \includegraphics[width=9cm]{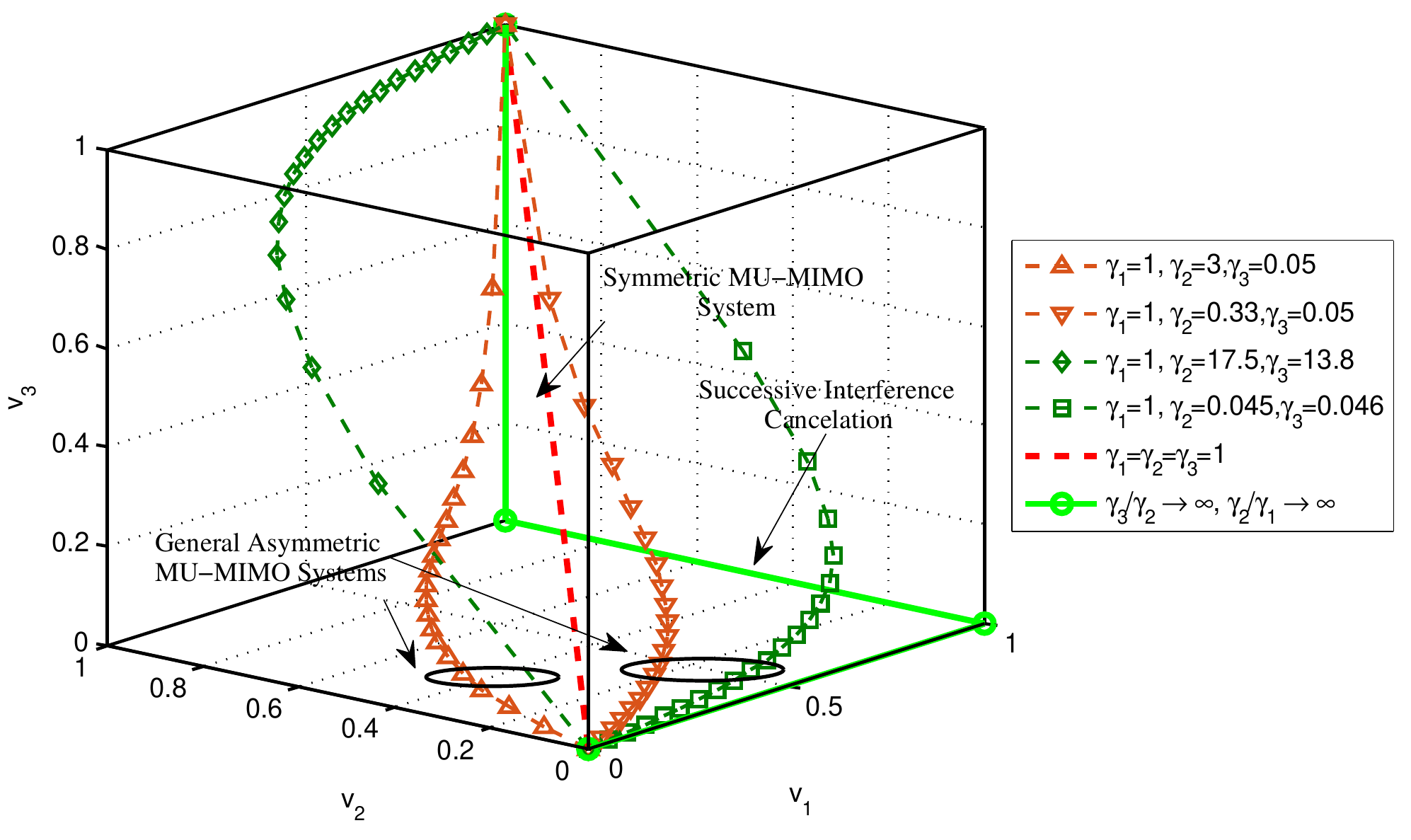}\\\vspace{-0.2cm}
  \caption{Variance tracks for the different $\bm{\gamma}$, where $\gamma_1=1$ is fixed. The variance of user $i$ is denoted as $v_i$, $i=1,2,3$. The variance track changes with $\gamma_2$ and $\gamma_3$. When $\gamma_3/\gamma_2\to\infty$ and $\gamma_2/\gamma_1\to\infty$ (green curve),  it is degenerated to a SIC case with the decoding order: user $3\to$user $2\to$ user $1$. When $\gamma_1=\gamma_2=\gamma_3=1$, it is degenerated to the symmetric case (red line). The other curves are the general asymmetric cases.}\label{f4}\vspace{-0.2cm}
\end{figure}

The area theorem tells us the achievable rate of every user is maximized if and only if its transfer function matches with that of the estimator and codes with that transfer function are existent. Therefore, we can arbitrarily choose the input variances of the estimator from the decoders and get users' achievable rate by matching the decoders' transfer functions with the estimator. To simplify to calculation, we let the input variances of the estimator satisfy the following constraints.
\begin{equation}\label{e45}
\gamma_i (v_i^{-1}-1)=\gamma_j (v_j^{-1}-1), \;\; \mathrm{for \;\; any} \;\;i,j\in \mathcal{N}_u.
\end{equation}
Without loss of generality, we assume $\gamma_1=1$ and $\gamma_i>0$ , that is, $v_i^{-1}=1+\gamma_i^{-1}(v_1^{-1}-1)$ for any $i\in \mathcal{N}_u$. Actually, the different $\bm{\gamma} = [\gamma_1, \cdots, \gamma_{N_u}]$ values give the different variance track during the iteration. Fig. \ref{f3} and Fig. \ref{f4} presents the variance tracks of the different $\bm{\gamma}$ for the two users and three users cases respectively. As we can see, when (\ref{e45}) concluded the symmetric case (when $w_1=\cdots =w_{N_u}$) and all the SIC points (maximal extreme points of the capacity region). If $\gamma_{k_i}/\gamma_{k_{i-1}}\to \infty$, for any $i\in \mathcal{N}_u/\{1\}$, we can get the SIC points with the decoding order $[k1,k2,\cdots,k_{N_u}]$, which is a permutation of $[1,2,\cdots,N_u]$. The blue curve and green curves in Fig. \ref{f3} and Fig. \ref{f4} are corresponding to some of the maximal extreme points. We will also show that the user's achievable rate can be adjusted by the parameter $\bm{\gamma}$.


With (\ref{e45}), we have
\begin{equation}\label{e46}
\mathbf{V}_{\bar{\mathbf{x}}}^{-1}= \mathbf{I}_{N_u} + \gamma_i(v_i^{-1}-1)\bm{\Lambda}_{\bm{\gamma}}^{-1} = \mathbf{V}_{\bar{\mathbf{x}}}^{-1}(v_i)
\end{equation}
and
\begin{eqnarray}\label{evall}
\mathbf{V}_{\hat{\mathbf{x}}} = (\sigma _{{{n}}}^{- 2}\mathbf{H}'^H\mathbf{H}'+ \mathbf{V}_{\bar{\mathbf{x}}}^{-1}(v_i))^{-1}
= \mathbf{V}_{\hat{\mathbf{x}}}(v_i)
\end{eqnarray}
for any $i\in\mathcal{N}_u$, where $\bm{\Lambda}_{\bm{\gamma}}=\mathrm{diag}(\bm{\gamma})$ is a diagonal matrix whose diagonal entries are $\bm{\gamma}$. Thus, we have
\begin{equation}\label{ephi}
\phi_i(\mathbf{v}_{\bar{\mathbf{x}}})= v_{\hat{x}_i}(v_i)^{-1}-v_i^{-1}=\phi_i(v_i)=\rho_i,
\end{equation}
For example, if we take $i=1$, we have
\begin{equation}\label{ev_1}
\mathbf{V}_{\bar{\mathbf{x}}}^{-1}= \mathbf{V}_{\bar{\mathbf{x}}}^{-1}(v_1),\;\;
\mathbf{V}_{\hat{\mathbf{x}}} =  \mathbf{V}_{\hat{\mathbf{x}}}(v_1),\;\;
\phi_i(\mathbf{v}_{\bar{\mathbf{x}}})= \phi_i(v_1).
\end{equation}

\emph{ Proposition 5: Based on (\ref{ephi}), for any $i\in \mathcal{N}_u$, the matching condition (\ref{e29}) can be rewritten as}
\begin{eqnarray}
\psi_i(\rho_i)&=&\phi_i^{-1}(\phi_i({1}))=1, \;\;\mathrm{for}\;\; 0\leq\rho_i<\phi_i({1});\label{em1}\\
\psi_i(\rho_i)&=&\phi_i^{-1}(\rho_i),\;\;\mathrm{for}\;\; \phi_i({1})\leq\rho_i<\phi_i({0});\label{em2}\\
\psi_i(\rho_i)&=&0,\;\;\mathrm{for}\;\; \phi_i({0})\leq\rho_i<\infty.\label{em3}
\end{eqnarray}

Then, we can give users' achievable rates of the iterative LMMSE detection for the MIMO-NOMA systems.

\emph{\textbf{Lemma 1}: For the MIMO-NOMA systems with iterative LMMSE detection, the achievable rates of the users are
\vspace{-0.2cm}
\begin{equation}\label{lemma1}
R_i = \int\limits_{v_1=1}^{v_1=0} \left[v_1- \gamma_i^{-1} \left[\mathbf{V}_{\hat{\mathbf{x}}}(v_1)\right]_{i,i} \right] dv_1^{-1}
- \log(\gamma_i),
\vspace{-0.2cm}
\end{equation}
where $\mathbf{V}_{\hat{\mathbf{x}}}(v_1)=\left( \sigma^{-2}_n\mathbf{H}'^H\mathbf{H}'+ \mathbf{I}_{N_u} + (v_1^{-1}-1)\bm{\Lambda}_{\bm{\gamma}}^{-1}\right)^{-1}$ and $[\cdot]_{i,i}$ denotes the $i$-th column and $i$-th row entry of the corresponding matrix.}

\begin{proof}
The achievable rate of user $i$ can be given by (\ref{e35}) with the matching condition.
\begin{eqnarray}\label{e47}
R_i\!\!\!\!\!&=&\int\limits_0^\infty  {{{\left( {{\rho _i} + {\psi _i}{{({\rho _i})}^{ - 1}}} \right)}^{ - 1}}d{\rho _i}}\nonumber\\
&\mathop \leq \limits^{(a)}&\!\!\!\int\limits_{\phi_i(1)}^{\phi_i(0)}  {{\left[ {\rho_i } +  \left( {\phi_i }^{ - 1}{({\rho_i })}\right)^{ - 1} \right]}^{ - 1}d{\rho_i }} +\!\!\! \int\limits_{0}^{\phi_i(1)} {(1+\rho_i)^{-1}d\rho_i}  \nonumber\\
&\mathop = \limits^{(b)}&\int\limits_{v_i=1}^{v_i=0}  {\left( v_i^{-1} +  \phi_i(v_i) \right)^{ - 1}d{\phi_i(v_i) }} + \log\left(1+\phi_i(v_i)\right)  \nonumber\\
&\mathop = \limits^{(c)}&\!\!\!\!\!\!\!\int\limits_{v_i=1}^{v_i=0} \!\!\! { v_{\hat{x}_i}(v_i) d {v_{\hat{x}_i}(v_i)}^{-1}} \!\!-\!\!\!\! \int\limits_{v_i=1}^{v_i=0} \!\!\!\! {v_{\hat{x}_i}(v_i)d{v_i^{-1} }} \!\!- \log v_{\hat{x}_i}(v_i=1)  \nonumber\\
&\mathop = \limits^{(d)}& - \int\limits_{v_1=1}^{v_1=0} { \gamma_i^{-1} \left[\mathbf{V}_{\hat{\mathbf{x}}}(v_1)\right]_{i,i} dv_1^{-1}}
-\mathop {\lim }\limits_{v_1 \to 0} \; \log\left[\mathbf{V}_{\hat{\mathbf{x}}}(v_1)\right]_{i,i}\;\nonumber\\
&\mathop = \limits^{(e)}&\!\!\! \int\limits_{v_1=1}^{v_1=0} \left[v_1- \gamma_i^{-1} \left[\mathbf{V}_{\hat{\mathbf{x}}}(v_1)\right]_{i,i} \right] dv_1^{-1} - \log(\gamma_i).
\end{eqnarray}
The inequality $(a)$ is derived by (\ref{em1})$\sim$(\ref{em3}) and the equality holds if and only if there exists that code whose transfer function satisfies the matching condition. The equations $(b)\sim(d)$ are given by $\rho_i=\phi_i(v_i)$, (\ref{ephi}) and (\ref{ev_1}), equation $(e)$ comes from  (\ref{e46}) and (\ref{evall}).

\begin{figure}[h]
  \centering
  \includegraphics[width=8cm]{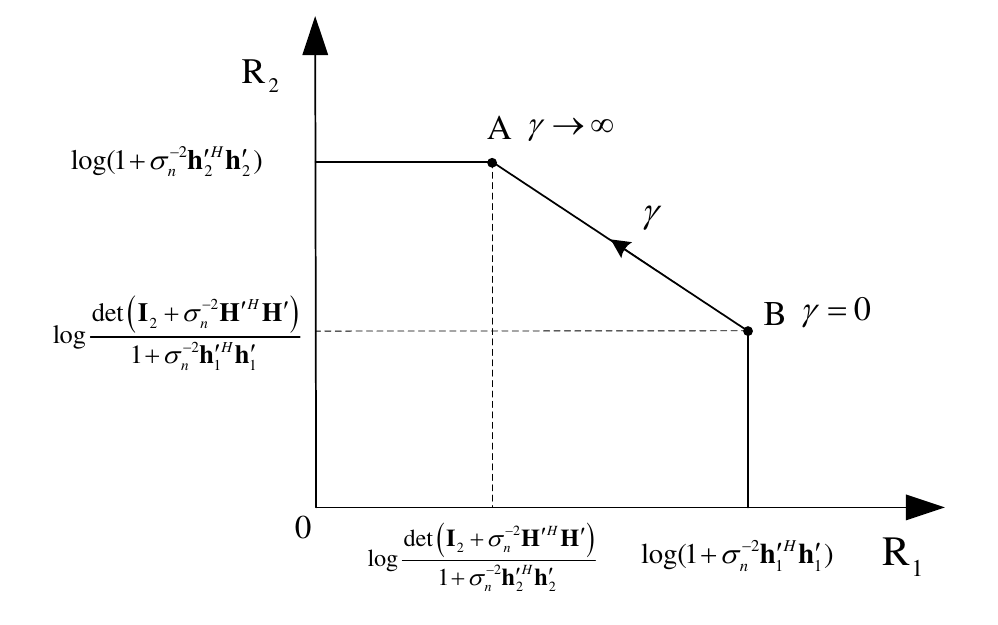}\\\vspace{-0.4cm}
  \caption{Capacity region achieving of iterative LMMSE detection for two-user MIMO system. When the parameter $\gamma$ changes from $0$ to $\infty$, the point $(R_1,R_2)$ moves from maximal extreme point B to maximal extreme point A along the segment AB. }\label{f5}
\end{figure}
\begin{figure}[h]
  \centering
  \includegraphics[width=7.7cm]{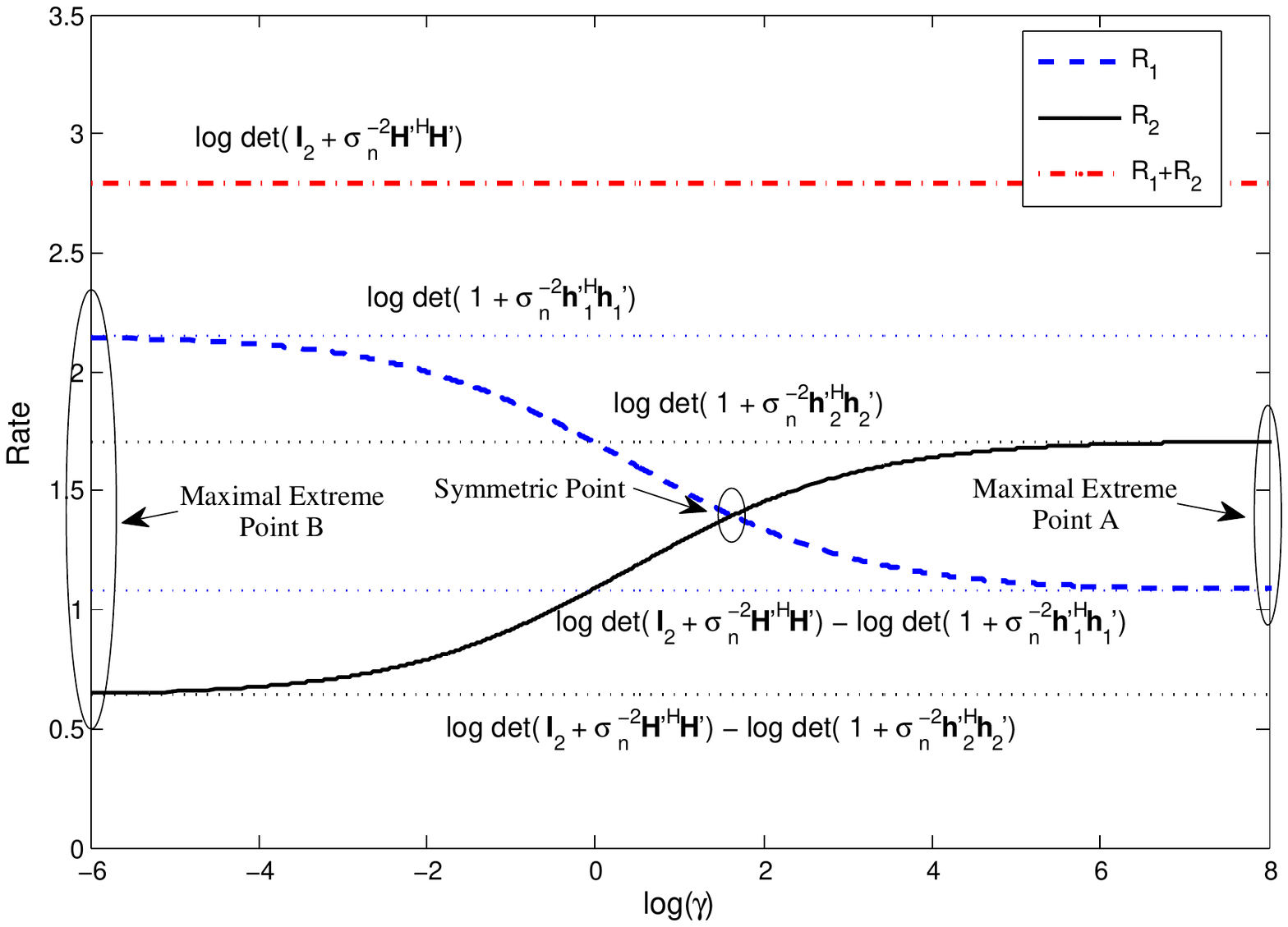}\\\vspace{-0.3cm}
  \caption{Relationship between the user rates and parameter $\gamma$ of the iterative LMMSE detection for two-user MIMO system. $N_r=2$, $\sigma_N^2=0.5$ and $\mathbf{H}=[
  1.32\;-1.31; \; -1.43 \;0.74]$. }\label{f6}\vspace{-0.2cm}
\end{figure}

Next, we show the existence of such codes whose transfer functions match the transfer functions of the LMMSE estimator. From the ``\emph{Property of SCM Codes}", in order to show the existence of such code, it only needs to check the matched transfer function meets the conditions (i)$\sim$(iv). It is easy to see that conditions (i) and (iv) are always satisfied by (\ref{em1}) and (\ref{em2}) respectively. From (\ref{ephi})$\sim$(\ref{em3}), we can see that $\psi_i(\rho_i)$ is continuous and differentiable in $[0,\infty)$ except at $\rho_i=\phi_i(0)$ and $\rho_i=\phi_i(1)$. Thus, Condition (iii) is satisfied. To show the monotonicity of the transfer function, we first rewrite (\ref{ese1}) as
\begin{eqnarray}\label{er2}
\phi_i(v_i)
 =  {{1 \mathord{\left/
 {\vphantom {1 {\left( {f_i^{ - 1}(v_i) - 1} \right)}}} \right.
 \kern-\nulldelimiterspace} {\left( {f_i^{ - 1}(v_i) - 1} \right)}}},
\end{eqnarray}
where $f_i(v_i)={\frac{{{w^2}}}{{\sigma _n^2}}\mathbf{h}_i^H{{\left( {{v_i^{ - 1}}{{\bf{I}}_{{N_r}}} + \frac{{{w^2}}}{{\sigma _n^2}}{\bf{H}}{{\bf{H}}^H}} \right)}^{ - 1}}{\mathbf{h}_i}}$. As $f_i(v_i)$ is a decreasing function of $v_i$, $\phi_i(v_i)$ is a decreasing function of $v$. With the definition of $\psi(\rho)$ from (\ref{em1})$\sim$(\ref{em3}), we can see that $\psi_i(\rho_i)$ is a monotonically decreasing function in $[0,\infty)$. Therefore, the matched transfer function can be constructed by the SCM code, i.e., there exists such codes that satisfy the matching condition.
\end{proof}

It should be noted that although the \emph{Lemma 1} gives the achievable rate of the users with respect to $\Lambda_{\bm{\gamma}}$, we cannot see the specific relationships between the achievable rates and $\Lambda_{\bm{\gamma}}$ because of the complicated integral structure of (\ref{lemma1}). Therefore, it is very hard to analyse the achievable rate region of the iterative LMMSE detection. However, its sum rate can be shown by the following theorem.

\emph{\textbf{Theorem 1:} The iterative LMMSE detection achieves the sum capacity of the MU-MIMO systems, i.e., $R_{sum}=\log \det\left(I_{N_u} + {\sigma_n^{-2}}\mathbf{H}'\mathbf{H}'^H\right)$.}

\begin{IEEEproof}
With (\ref{lemma1}), the achievable sum rate is\vspace{-0.2cm}
\begin{eqnarray}\label{e49}
&\!\!\!\!\!\! R_{sum}&\!\!\!\!=\sum\limits_{i = 1}^{{N_u}} {{R_i}} \nonumber \\
 &\!\!\!\!\!\!\!\!\!\!\!\!\!\!\!\!\mathop\leq \limits^{(a)}&\!\!\!\!\!\!\!\!\!\!\!\!\!-\!\!\!\! \int\limits_{v_1=1}^{v_1=0} \!\sum\limits_{i = 1}^{{N_u}} {{ \left(\gamma_i^{-1} \left[\mathbf{V}_{\hat{\mathbf{x}}}(v_1)\right]_{i,i} \right) dv_1^{-1}}  }\!\!
-\!\!\!\mathop {\lim }\limits_{v_1 \to 0} \; \log(v_1^{N_u}\mathop \Pi \limits_{i = 1}^{{N_u}} \gamma_i) \nonumber\\
 &\!\!\!\!\!\!\!\!\!\!\!\!\!\!\!\!\!\!=&\!\!\!\!\!\!\!\!\!\!\!\!\!\!- \!\!\int\limits_{v_1=1}^{v_1=0}  {{ \mathrm{Tr}\{\bm{\Lambda}_{\bm{\gamma}}^{-1} \mathbf{V}_{\hat{\mathbf{x}}}(v_1) \} dv_1^{-1}}  }
-\mathop {\lim }\limits_{v_1 \to 0} \; \log(v_1^{N_u}\mathop \Pi \limits_{i = 1}^{{N_u}} \gamma_i) \nonumber\\
 &\!\!\!\!\!\!\!\!\!\!\!\!\!\!\!\!\!\!\mathop = \limits^{(b)}&\!\!\!\!\!\!\!\!\!\!\!\!\!\!\!\!\!-\!\!\!\mathop {\lim }\limits_{v_1 \to 0} \!\! \log(v_1^{N_u}\!\!\!\mathop \Pi \limits_{i = 1}^{{N_u}}\!\!\! \gamma_i\!)\!\!-\!\! \left[\! \log\!\det\!\left( \!(v_1^{-1}\!\!\!\!-\!1\!)\mathbf{I}_{N_u}\!\! \!\!+\!\! \left(\! \mathbf{I}_{N_u}\! \!\!\!+\! \sigma_n^{-2}\mathbf{H}'^H\!\mathbf{H}' \!\right)\!\!\bm{\Lambda}_{\bm{\gamma}} \!\right)\!\right]_{\!v_1\!=\!1}^{\!v_1\!=\!0}
 \nonumber\\
  &\!\!\!\!\!\!\!\!\!\!\!\!\!\!\!\!\!\!=&\!\!\!\!\!\!\!\!\!\!\!\!\!\!\!\!\!-\!\!\!\mathop {\lim }\limits_{v_1 \to 0} \!\! \log(v_1^{N_u}\!\!\!\mathop \Pi \limits_{i = 1}^{{N_u}} \!\!\!\gamma_i\!)\!\!-\!\!\! \mathop {\lim }\limits_{v_1 \to 0}\!\! {\log \! \det(v_1^{-1}\!\mathbf{I}_{N_u}\!) }\!\! +\!\! \log \!\det\!\left(\! \left(\!\mathbf{I}_{N_u} \!\!\!\!+ \!\!{\sigma _n^{- 2}}\!\mathbf{H}'^H\!\mathbf{H}'\!\right)\!\!\bm{\Lambda}_{\bm{\gamma}}\! \right)
 \nonumber\\
&\!\!\!\!\!\!\!\!\!\!\!\!\!\!\!\!\!\!=& \!\!\!\!\!\!\!\!\!\!\!\!\!\log \det\left( \mathbf{I}_{N_u} + {\sigma _n^{- 2}}\mathbf{H}'^H\mathbf{H}' \right)
\end{eqnarray}
which is the exact sum capacity of the system. The inequality $(a)$ is derived by (\ref{e49}), and equation $(b)$ is based on (\ref{evall}) and the law $\int {\mathrm{Tr}\{ \left(s\mathbf{I} + \mathbf{A}\right)^{-1}\} ds }= \log\det (s\mathbf{I} + \mathbf{A})$. It means that the iterative detector can achieve the system sum capacity with different kinds of user rate combinations.
\end{IEEEproof}

Theorem 2 shows that for a general MIMO-NOMA system, from the sum rate perspective, the iterative detection structure is optimal and the LMMSE estimator is an optimal estimator without losing any useful information during the estimation.

\section{Special Cases of the MIMO-NOMA Systems}
In the last section, we proved the iterative LMMSE detection achieves the sum capacity of the MU-MIMO systems, but whether it can achieve the whole capacity region is still unkown. In this section, we analyse some special cases of the MIMO-NOMA systems.

\subsection{ Maximal Extreme Points Achieving}
In this subsection, we analyse the maximal extreme points in the achievable rate region of the iterative LMMSE detection.

\textbf{\emph{Corollary 1}}: \emph{The maximal extreme points in MU-MIMO capacity region are achieved by the iterative LMMSE detection.}

\begin{IEEEproof}
The proof is omitted due to the page limit.
\end{IEEEproof}

These corollary shows that as the parameter $\Lambda_{\bm{\gamma}}$ be properly chosen, the iterative LMMSE detection can be degenerated to the SIC methods, i.e., the SIC methods are some special cases of the proposed iterative LMMSE detection.

\subsection{ Capacity Region Achieving for Two-user MIMO Systems}

As it is mentioned, it is hard to calculate the specific achievable rates of users for the general MIMO-NOMA systems. In this subsection, we show that the iterative LMMSE detection is capacity region achieving for two-user MIMO systems.

\emph{\textbf{Theorem 2}}: \emph{The iterative LMMSE detection achieves the whole capacity region of two-user MIMO systems as follows.}
\begin{equation}
\left\{ \begin{array}{l}
R_1\leq\log ( {1 + \frac{1}{{\sigma _n^2}}{\mathbf{h}'}_1^H{{\mathbf{h}'}_1}} ),\\
R_2\leq\log ( {1 + \frac{1}{{\sigma _n^2}}{\mathbf{h}'}_2^H{{\mathbf{h}'}_2}} ),\\
R_1+R_2\leq\log \det\left( \mathbf{I}_{2} + {\sigma _n^{- 2}}\mathbf{H}'^H \mathbf{H}'\right).
\end{array} \right.
\end{equation}

\begin{IEEEproof}
The proof is omitted due to the page limit.
\end{IEEEproof}

For two-user case, the $R_1$ and $R_2$ is given as follows.

\emph{\textbf{Corollary 2}:  The user rates of the proposed iterative LMMSE detection for two-user MIMO system are}
\begin{equation}\label{coro2_1}
\left\{ \begin{array}{l}
R_1=\frac{1}{2}\log(\gamma\det(A)) +\frac{a_{22}\gamma-a_{11}}{2\eta} \log\frac{a_{22}\gamma+a_{11}-\eta}{a_{22}\gamma+a_{11}+\eta},\\
R_2= \frac{1}{2}\log(\gamma^{-1}\det(A)) -\frac{a_{22}\gamma-a_{11}}{2\eta} \log\frac{a_{22}\gamma + a_{11}+\eta}{a_{22}\gamma+a_{11}+\eta}
\end{array} \right.
\end{equation}
where $\mathbf{A}=\sigma^{-2}_n\mathbf{H}'^H\mathbf{H}'+ \mathbf{I}_{2}=\left[  {\begin{array}{*{20}{c}}
  \vspace{-0.2cm} {{a_{11}}}&{{a_{12}}} \\
  {{a_{21}}}&{{a_{22}}}
\end{array}} \right]$ and $\eta=\sqrt {a_{22}^2{\gamma ^2} + 2(2{a_{21}}{a_{12}} - {a_{22}}{a_{11}})\gamma  + a_{11}^2} $. It is easy to find that $\eta$ is a real number as $\mathbf{A}$ is positive definite and $\gamma\geq0$.

\textbf{\emph{Remark 2:}} It should be noted from (\ref{coro2_1}) that $R_1$ and $R_2$ are not linear functions of $\gamma$. It is easy to check that $R_1+R_2=\log \det\left( \mathbf{I}_{2} + {\sigma _n^{- 2}}\mathbf{H}'^H \mathbf{H}'\right)$, and when $\gamma\to0$ (or $\gamma\to\infty$), the limit of $(R_1,R_2)$ in (\ref{coro2_1}) converges to the maximal point B (or A) in Fig. \ref{f5}. When the parameter $\gamma$ changes from $0$ to $\infty$, the point $(R_1,R_2)$ can achieve any point on the segment AB in Fig. \ref{f5}. It also shows another proof of \emph{Theorem 3}. Fig. \ref{f6} presents the rate curves of $R_1$ and $R_2$ respect to the parameter $\gamma$. It verifies that $R_2$ increases monotonously with the $\gamma$ and $R_1+R_2$ always equals to the system sum capacity.

\section{Conclusion}
An iterative LMMSE detector for the MIMO-NOMA systems has been studied, which has a low-complexity as the distributed processors replace the overall receiver. The achievable rate of the iterative LMMSE detector has been analysed, which shows that the iterative LMMSE detector is sum capacity achieving for the MU-MIMO systems. In addition, we proved that with the carefully designed iterative LMMSE detector, all the maximal extreme points in the capacity region of MU-MIMO systems are achievable, and the whole capacity region of two-user MIMO systems are also achievable.



\begin{thebibliography}{h}
\bibitem{5GWhitepaper}
 ``5G radio access: requirements, concepts and technologies," \emph{NTT DOCOMO, Inc., Tokyo, Japan, 5G Whitepaper}, Jul. 2014.
\bibitem{Yang2014}
J. Yang, S. Xie, X. Zhou, R. Yu, Y. Zhang, ``A Semiblind Two-Way Training Method for Discriminatory Channel Estimation in MIMO Systems", \emph{IEEE Trans. on Commun.}, vol. 62, no.7, pp.2400-2410, July 2014.
\bibitem{Argas2013}
D. Argas, D. Gozalvez, D. Gomez-Barquero, and N. Cardona, ``MIMO for DVB-NGH, the next generation mobile TV broadcasting," \emph{IEEE Commun. Mag.}, vol. 51, no. 7, pp. 130-137, Jul. 2013.
\bibitem{Rusek2013}
F. Rusek, D. Persson, B. K. Lau, E. G. Larsson, T. L. Marzetta, O. Edfors, and F. Tufvesson, ``Scaling up MIMO: Opportunities and challenges with very large arrays," \emph{IEEE Signal Process. Mag.}, vol. 30, no. 1, pp. 40-60, Jan. 2013.
\bibitem{biglieri2007}
E. Biglieri, R. Calderbank, A. Constantinides, A. Goldsmith, A. Paulraj, and H. V. Poor, \emph{MIMO Wireless Communications}. Cambridge University Press, Cambridge, 2007.

\bibitem{Ngo2012}
H. Ngo, E. Larsson, and T. Marzetta, ``Energy and spectral efficiency of very large multiuser MIMO systems," \emph{IEEE Trans. Commun.}, vol. 61, no. 4, pp. 1436-1449, Apr. 2012.
\bibitem{Dai2013}
L. Dai, Z. Wang, and Z. Yang, ``Spectrally efficient time-frequency training OFDM for mobile large-scale MIMO systems," \emph{IEEE J. Sel. Areas Commun.}, vol. 31, no. 2, pp. 251-263, Feb. 2013.
\bibitem{Han_acpt}
Y. Han, S. Jin, X. Li, H. Zhang, R. Yu and Y. Zhang, ``Investigation of Transmission Schemes for Millimeter-Wave Massive MU-MIMO Systems", \emph{accepted by IEEE Systems Journal}.

\bibitem{Saito2013}
Y. Saito, Y. Kishiyama, A. Benjebbour, T. Nakamura, A. Li, and K. Higuchi, ``Non-orthogonal multiple access (NOMA) for cellular future radio access," in \emph{Proc. IEEE Vehicular Technology Conference, Dresden, Germany}, Jun. 2013.
\bibitem{Al-Imari2014}
M. Al-Imari, P. Xiao, M. A. Imran, and R. Tafazolli, ``Uplink non-orthogonal multiple access for 5g wireless networks," in \emph{Proc. of the 11th International Symposium on Wireless Communications Systems (ISWCS), Barcelona, Spain}, Aug 2014, pp. 781-785.
\bibitem{Ding2014}
Z. Ding, Z. Yang, P. Fan, and H. V. Poor, ``On the performance of non-orthogonal multiple access in 5G systems with randomly deployed users," \emph{IEEE Signal Process. Letters}, vol. 21, no. 12, pp. 1501-1505, Dec 2014.
\bibitem{Lei1}
L. Liu, Y. Li, Y. Su and Y. Sun, ``Quantize-and-Forward Strategy for Interleave Division Multiple-Access Relay Channel," \emph{accepted by IEEE Transactions on Vehicular Technology}, 2015.
\bibitem{Lei2}
L. Liu, Y. Li, Y. Chau, Y. L. Guan and Y. Sun, ``Distributed Joint Source-Channel Superposition Coding for Asymmetric Correlation-Actual Channel," \emph{submitted to IEEE Transactions on Vehicular Technology}, 2016.

\bibitem{Micciancio2001}
D. Micciancio, ``The hardness of the closest vector problem with
preprocessing," \emph{IEEE Transactions on Information Theory}, vol. 47, no. 3, pp. 1212-1215, Mar. 2001.
\bibitem{verdu1984_1}
S. Verd\'{u}, ``Optimum multi-user signal detection," Ph.D. dissertation, Department of Electrical and Computer Engineering, University of Illinois at Urbana-Champaign, Urbana, IL, Aug. 1984.
\bibitem{verdu1987}
S. Verd\'{u} and H. V. Poor, ``Abstract dynamic programming models under commutativity conditions," \emph{SIAM Journal on Control and Optimization}, vol. 25, no. 4, pp. 990-1006, Jul. 1987.


\bibitem{tse2005}
Tse David and Pramod Viswanath, \emph{Fundamentals of wireless communication.} Cambridge university press, 2005.
\bibitem{Loeliger2004}
H. A. Loeliger, ``An introduction to factor graphs," \emph{IEEE Signal Processing Mag.}, pp. 28-41, Jan. 2004.
\bibitem{Loeliger2006}
H. A. Loeliger, J. Hu, S. Korl, Q. Guo and L. Ping, ``Gaussian message passing on linear models: an update," \emph{Int. Symp. on Turbo codes and Related Topics}, Apr. 2006.
\bibitem{andrea2005}
A. Montanari, B. Prabhakar, and David Tse, ``Belief Propagation Based Multi-User Detection," \emph{Proceedings}, Vol. 43, 2005.

\bibitem{Gao2014}
X. Gao, L. Dai, C. Yuen, and Y. Zhang, ``Low-Complexity MMSE Signal Detection Based on Richardson Method for Large-Scale MIMO Systems," \emph{in IEEE 80th Vehicular Tech. Conf. (VTC Fall)}, Sept. 2014, pp. 1-5.
\bibitem{Lei2015}
L. Liu, Y. Chau, Y. L. Guan, Y. Li and Y. Su, ``A Low-Complexity Gaussian Message Passing Iterative Detection for Massive MU-MIMO Systems," in \emph{Proc IEEE International Conference on Information, Communications and Signal Processing (ICICS)}, Singpore, Dec. 2015.
\bibitem{verdu1998}
S. Verd\'{u}, \emph{Multiuser Detection}. Cambridge, UK: Cambridge University Press, 1998.

\bibitem{Ping2003_1}
L. Ping, L. Liu, K. Y. Wu, and W. K. Leung, ``Interleave-division multiple-access (IDMA) communications," in \emph{Proc. Int. Symp. Turbo Codes Related Topics, Brest, France, Sept. 2003}, pp. 173-180.
\bibitem{Guo2008}
Q. Guo and L. Ping, ``LMMSE turbo equalization based on factor graphs," \emph{IEEE J. Sel. Areas Commun.}, vol. 26, no. 2, pp. 311-319, 2008.
\bibitem{Caire2004}
G. Caire, R. Muller, and T. Tanaka, ``Iterative multiuser joint decoding: Optimal power allocation and low-complexity implementation," \emph{IEEE Trans. Inf. Theory}, vol. 50, no. 9, pp. 1950-1973, Sep. 2004.
\bibitem{Yuan2014}
X. Yuan, L. Ping, C. Xu and A. Kavcic, ``Achievable Rates of MIMO Systems With Linear Precoding and Iterative LMMSE Detection," \emph{IEEE Trans. Inf. Theory}, vol. 60, no.11, pp. 7073-7089, Oct. 2014.
\bibitem{Bhattad2007}
K. Bhattad and K. R. Narayanan, ``An MSE-based transfer chart for
analyzing iterative decoding schemes using a Gaussian approximation," \emph{IEEE Trans. Inf. Theory}, vol. 53, no. 1, pp. 22-38, Jan. 2007.
\bibitem{Guo2005}
D. Guo, S. Shamai, and S. Verd\'{u}, ``Mutual information and minimum
mean-square error in Gaussian channels," \emph{IEEE Trans. Inf. Theory}, vol. 51, no. 4, pp. 1261-1282, Apr. 2005.
\bibitem{Ashikhmin2004}
A. Ashikhmin, G. Kramer, and S. ten Brink, ``Extrinsic information
transfer functions: Model and erasure channel properties," \emph{IEEE Trans. Inf. Theory}, vol. 50, no. 11, pp. 2657-2673, Nov. 2004.
\bibitem{Brink2001}
S. ten Brink, ``Convergence behavior of iteratively decoded parallel concatenated codes," \emph{IEEE Trans. Commun.}, vol. 49, no. 10, pp. 1727-1737, Oct. 2001.
\bibitem{Kay1993}
S. Kay, \emph{Fundamentals of Statistical Signal Processing: Estimation Theory.} Upper Saddle River, NJ, USA: Prentice-Hall, 1993.
\bibitem{Poor1997}
H. V. Poor and S. Verd\'{u}, ``Probability of error in MMSE multiuser detection," \emph{IEEE Trans. Inf. Theory}, vol. IT-43, no. 3, pp. 835-847, May 1997.
\bibitem{Wachsmann1999}
U. Wachsmann, R. F. H. Fischer, and J. B. Huber, ``Multilevel codes: Theoretical concepts and practical design rules," \emph{IEEE Trans. Inf. Theory}, vol. 45, no. 5, pp. 1361-1391, Jul. 1999.
\bibitem{Gadkari1999}
S. Gadkari and K. Rose, ``Time-division versus superposition coded modulation schemes for unequal error protection," \emph{IEEE Trans. Commun.}, vol. 47, no. 3, pp. 370-379, Mar. 1999.
\end{thebibliography}
\end{document}